\documentclass[aps,pre,twocolumn,groupedaddress,showpacs,floatfix]{revtex4}
\usepackage{graphicx}
\usepackage{amsmath}
\usepackage{graphicx}
\usepackage{graphics}
\usepackage{amsfonts}
\usepackage{amssymb}

\usepackage[utf8]{inputenc}

\begin{document}

\title{Rate Equations and Scaling in Pulsed Laser Deposition}

\author{A. C. Barato,$^1$ H. Hinrichsen,$^1$ and D. E. Wolf$^2$}

\affiliation{$^1$Fakult{\"a}t f{\"u}r Physik und Astronomie, Universit{\"a}t W{\"u}rzburg, Am Hubland, 97074 W{\"u}rzburg, Germany\\
	$^2$Dept. of Physics, University Duisburg-Essen, Campus Duisburg, D-47048 Duisburg, Germany}

%-------------------------------------------------------
\begin{abstract}
We study a simplified model for pulsed laser deposition [Phys. Rev. Lett. {\bf 87}, 135701 (2001)] by rate equations. We consider a set of equations, where islands are assumed to be point-like, as well as an improved one that takes the size of the islands into account. The first set of equations is solved exactly but its predictive power is restricted to a few pulses. The improved set of equations is integrated  numerically, is in excellent agreement with simulations, and fully accounts for the crossover from continuous to pulsed deposition. Moreover, we analyze the scaling of the nucleation density and show numerical results indicating that a previously observed logarithmic scaling does not apply. 
\end{abstract}
%-------------------------------------------------------

% Explanation of PACS numbers:
% 68.43.Jk: Diffusion of adsorbates, kinetics of coarsening and aggregation
% 68.55.A-: Nucleation and growth
% 05.45.Df: Fractals

\maketitle
\parskip 1mm 

%-------------------------------------------------------
\section{Introduction}
%-------------------------------------------------------

Among the techniques for the growth of thin films and multilayers Molecular Beam Epitaxy (MBE) and Pulsed Laser Deposition (PLD)~\cite{chri94} play an important role. Compared to MBE, PLD has several advantages. For example, pulsed laser deposition is a widely used technique which allows one to control the stoichiometry of multilayers more efficiently and leads in some situations to a better layer-by-layer growth~\cite{shen04}. The essential difference between these two techniques lies in the way how particles are deposited and the kinetic energy of the arriving particles. While in MBE the flux of particles is continuous, in PLD intense and short pulses are deposited on the target, making the flux of incoming particles approximately discontinuous. Also, in PLD the kinetic energy of the deposited particles is typically higher than in MBE.

There is already a great amount of theoretical work on MBE~\cite{bara95}, while there are less theoretical studies on PLD~\cite{jens97, hinn01, lam02, lee03, vasc06, vasc07, vasc08}. In the present paper we apply to PLD a theoretical method that proved to be very useful in MBE, namely, the rate equations approach~\cite{vena73}. By integrating a set of rate equations for the island density one can make predictions about the experimentally relevant quantities characterizing the system. We use two sets of rate equations: In the first and more simple one, we consider the islands as point-like objects, while in the second approach, introduced for MBE in Ref.~\cite{bale94}, we improve the equations by considering the competition of islands of different sizes for the arriving and diffusing monomers. The first approach has the advantage that the rate equations can be solved exactly. The improved set of rate equations can only be integrated numerically but it shows excellent agreement with simulations.

In this work we study PLD  on the basis of a model investigated in Ref.~\cite{hinn01}, where monomers are deposited on the surface in infinitely short pulses. These monomers diffuse on the surface between subsequent pulses until they nucleate to form the seeds of immobile islands. More precisely, the model is defined as follows:
\begin{itemize}
\item[-] In each pulse $I\cdot L^2$ monomers, where $I$ is the pulse intensity, are randomly deposited on a two-dimensional $L\times L$ lattice.
\item[-] During the time interval between two pulses of length $\tau= I/F$, where $F$ is the time-averaged influx of particles, the monomers diffuse with rate $D$.
\item[-] When a monomer encounters another monomer or the border of an agglomerate of particles, it sticks irreversibly.
\end{itemize}
The model does not take into account that the arriving particles in PLD have a high kinetic energy, leading to transient effects such as an enhanced mobility. In fact, the transient mobility was shown to have a great influence on the growth kinetics \cite{vasc06, vasc07, vasc08}. However, Vasco et al. \cite{vasc08}, analyzing a set of rate equations, found out that only the way particles are deposited already leads to differences in the film roughness. More specifically, they found that continuous deposition leads generally to rougher films and an additional transient mobility in the case of continuous (discontinuous) deposition makes the film rougher (smoother). Other simplifications assumed in the model are the irreversibility of aggregation and the immobility of nucleated islands. For instance, the assumption of irreversible aggregation hinders to study Ostwald ripening, which was recently proposed to be the key mechanism ruling the PLD growth kinetics in momoner-depeted regimes \cite{vasc08}. Here we want to study the difference in the island morphology for continuous and discontinuous deposition and the corresponding crossover in the limit of low energies, assuming that transient mobility effects lead to corrections that can be neglected. Moreover, we restrict ourselves to the submonolayer regime, where the coverage $\theta$, which is given by the pulse intensity multiplied by the number of pulses, is less than one. It turns out that the dynamical processes at the bottom layer determine the growth process at the following layers to a large extent~\cite{brun98}. 

A nucleation event occurs when a diffusing monomer encounters another monomer. A quantity of interest in the study of PLD and MBE is the nucleation density $n$, which is defined as the number of nucleation events divided by the number of lattice sites. In Ref.~\cite{hinn01} it was suggested that for PLD this quantity follows an unusual logarithmic scaling. Later this type of scaling behavior was explained in the framework of scaling laws with continuously varying exponents~\cite{sitt02}. Opposed to these early conjectures, we present numerical evidence that such logarithmic scaling laws do not hold asymptotically, although they may be used as a good approximation.

The paper is organized in the following way. In the next section we present the properties of the model and introduce the observables that we are going to calculate. The third section is dedicated to the rate equations, where islands are approximated as point-like objects. In section 4 we introduce and analyze the improved set of rate equations. Section 5 is concerned with a critical analysis of scaling laws for PLD. Finally, our conclusions are summarized in section 6.

%-------------------------------------------------------
\section{Model properties}
%-------------------------------------------------------

Depending on the pulse intensity $I$ and on the ratio $R=D/F$ the model for PLD defined above displays different features. For very small pulse intensities and a finite value of $R$, it exhibits essentially the same behavior as MBE. More specifically, for intensities much smaller than a typical value $I_c$, which depends on $R$, the model is in the MBE regime. Then, increasing the pulse intensity to a value much larger than $I_c$, the model crosses over to a different regime, the so-called PLD regime. The two regimes differ in so far as they are characterized by different surface morphologies, that is, the number and size of islands are distributed differently. However, as edge diffusion is not included in the model, the islands remind of fractal objects in both cases and their fractal dimension $d_f$ is approximately equal to the one of diffusion limited aggregation (DLA)~\cite{witt81, meak86}. \cite{footnote}

The two regimes are also different with respect to the scaling of the average distance between islands $l_D$. In the MBE limit $I\ll I_c$ the average distance between islands grows as~\cite{bara95}  
\begin{equation} 
l_D\sim R^\gamma,
\label{eqldm}
\end{equation}
where 
\begin{equation} 
\gamma= \frac{1}{d_f+4}\,,
\label{eqgama}
\end{equation} 
while in the PLD limit $I\gg I_c$ one has~\cite{hinn01}
\begin{equation}
l_D\sim I^{-\nu},
\label{eqldp}
\end{equation}   
where 
\begin{equation}
\nu= \gamma/(1-2\gamma).
\label{eqnu}
\end{equation}
Hence, the crossover takes place if
\begin{equation}
I_c\approx R^{-\gamma/\nu}= R^{2\gamma-1}.
\end{equation}
Relations (\ref{eqldm}) and (\ref{eqldp}) can be combined in the following scaling form~\cite{hinn01}
\begin{equation}
l_D\sim R^{\gamma}\,h(I/R^{-\gamma/\nu}),
\label{eqcross}
\end{equation}  
where $h(x)$ is a scaling function which is constant for $x\ll 1$ and $h(x)\sim x^{-\nu}$ for $x\gg 1$. The same type of scaling behavior is also expected to hold when considering other typical length scales as, for example, the mean distance that a monomer travels before being captured by an island or another monomer, the square root of the mean island size, and the inverse of the square root of the nucleation density. In numerical simulations the scaling relations (\ref{eqldm}), (\ref{eqldp}) and (\ref{eqcross}) were observed to hold for small coverages before coalescence of islands sets in (see below). 

In the PLD limit the model is independent of $R$, the time interval between two pulses is big enough to make the density of monomers equal to zero. Hence, taking the PLD limit means that diffusion of monomers occurs until the density of them is zero, and in principle such a limit can be taken for any value of $I$. The smallest value of $I$ with which we perform calculations is $0.0001$. This value is much lower than typical experimental pulse intensities, which are of order $0.1$  \cite{stri98, vasc04}. We use low values for the pulse intensity because we are interested in the scaling behavior of the model in the limit $I\to 0$.

The probability that an occupied site belongs to a cluster of size $s$ is given by
\begin{equation}
p_s= \frac{sN_s}{\theta}\,,
\label{eqprob}
\end{equation}
where $N_s$ is the number-density of islands of size $s$ and 
\begin{equation}
\theta= Ft= \sum_{s=1}^{\infty}sN_s
\end{equation}
is the coverage. Therefore, the mean island size is given by
\begin{equation}
\langle s\rangle = \frac{1}{\theta}\sum_{s=1}^{\infty}s^2N_s.
\end{equation}
%
%=====================================================
\begin{figure}
\begin{center}
\includegraphics[width=85mm]{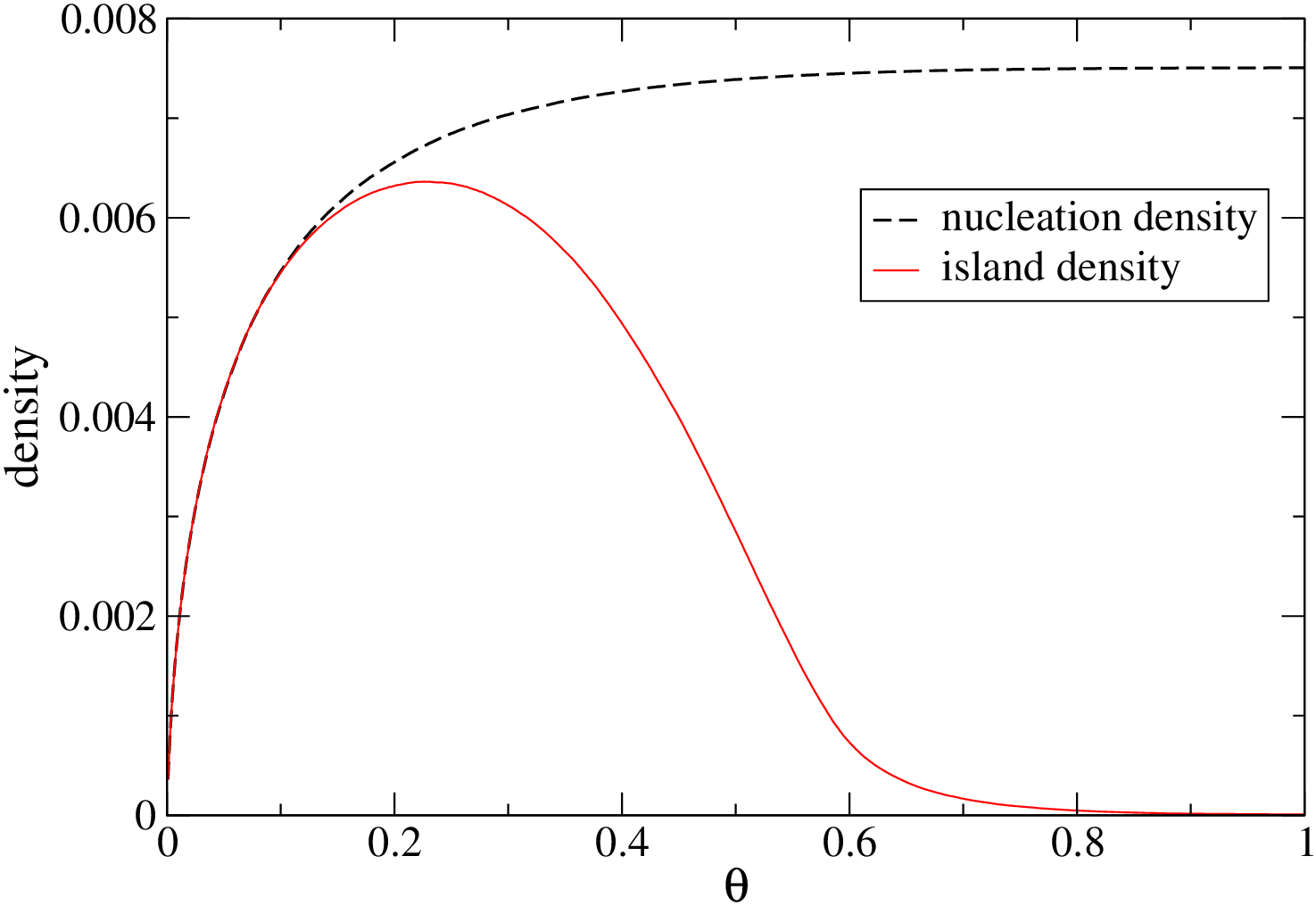}
\caption{(Color online) Nucleation density $n(t)$ and island density $N(t)$ obtained from simulations in the PLD limit for $I=0.001$.
\label{fnucid}}
\end{center}
\end{figure}
%=====================================================

\noindent
The nucleation density $n$ is defined as the number of nucleation events (when a monomer meets another monomer) in the bottom layer per unit area. Initially, when freshly nucleated islands do not yet touch each other, it is equal to the island density, given by
\begin{equation}
N= \sum_{s=2}^{\infty}N_s\,.
\label{eqisland}
\end{equation}
However, as the coverage increases, more and more islands coalesce, forming bigger islands. Consequently their number decreases so that the island density becomes smaller than the nucleation density, as is shown in Fig.\ref{fnucid}. As can be seen, coalescence -- indicated by a difference between $n$ and $N$ -- sets in already before the island density reaches its maximum. At this maximum the island density begins to decrease because coalescence events become more frequent than nucleation events.

As for the scaling of the nucleation density in the PLD regime, the following relations were observed in Ref.~\cite{hinn01}: After the first pulse, where $\theta=I$, the nucleation density scales as
\begin{equation}
n(I,I)\sim I\,.
\label{eqlin}
\end{equation}
Upon completion of the bottom layer ($\theta=1$) the nucleation density was found to scale as
\begin{equation}
n(I,1)\sim I^{2\nu}.
\label{eq1}
\end{equation}
These two scaling laws determine the terminal points of the curves $n(I,\theta)$ for different intensities $I$. However, these curves were found to have a different shape in a double-logarithmic representation, making it impossible to perform an ordinary data collapse. This suggested that the model does not exhibit ordinary power-law scaling.

As a possible way out, a logarithmic scaling form was proposed~\cite{hinn01}. Starting point was the observation that by squeezing and stretching the curves in such a way that the terminal points collapse, the entire curves for different intensities seem to collapse as well. This led to the conjectured scaling form
\begin{equation}
\ln M(I,\theta)= (\ln I)\,g\bigg(\frac{\ln\theta}{\ln I}\bigg),
\label{eqscalog}
\end{equation} 
where 
\begin{equation}
M(I,\theta)= \frac{n(I,\theta)}{n(I,1)}\,.
\label{eqM}
\end{equation}
Here $g(x)$ is a universal function, and this scaling is supposed to be valid in the full range $0<\theta\leq 1$. In a subsequent work, this type of scaling behavior was backed up by a theoretical framework which involves continuous varying exponents~\cite{sitt02}, where the scaling exponents are continuous functions of the control parameters. One of the purposes of this paper is to show that this logarithmic scaling does probably not apply to PLD in a strict sense.

Throughout this paper all numerical simulations were performed in 2+1 dimensions on a $400\times400$ lattice with periodic boundary conditions. For average quantities such as the island and nucleation densities or the mean island size, the number of independent runs was $100$ while, when calculating the probability distribution (\ref{eqprob}), we used $10000$ runs.  

%-------------------------------------------------------
\section{Rate equations}
%-------------------------------------------------------

In this and the following section we study the model introduced above by mean-field approximations on different levels. Before starting let us emphasize that these approaches are based on the same assumptions (e.g. irreversible aggregation, immobile aggregates) as the full model defined in the introduction, extended by additional mean-field type approximations such as the homogeneity of densities and the neglect of correlations.
 
We start with a simple set of rate equations for the island density $N$ and the monomer density $N_1$ (see Ref.~\cite{bara95, bale94, amar94})
\begin{equation}
\frac{dN_1}{dt}= F(t)- 2DN_1^2- DNN_1,
\label{eqN1}
\end{equation}
\begin{equation}
\frac{dN}{dt}= DN_1^2 \,.
\label{eqN}
\end{equation}
The first term on the right hand side of Eq. (\ref{eqN1}) describes the flux of particles while the second and third terms are related to nucleation of two monomers and attachment of monomers at existing islands, respectively. In Eq. (\ref{eqN}) the term on the right hand side describes the increase of the island density by nucleation of two diffusing monomers. For MBE the flux is constant, while for PLD the flux is modeled by a discontinuous sequence of spikes
\begin{equation}
F(t)= I\sum_{l=0}^\infty\delta(t-\tau l),
\end{equation}  
where $\tau$ is the time interval between two pulses. 

In this set of rate equations coalescence is not taken into account, hence within this theory the nucleation density and the island density are identical. Moreover, the approach does not take into account how islands of different sizes compete for the diffusing and arriving monomers since islands are treated as point-like objects. Therefore, this approximation is expected to be valid (when compared with simulations) just for the first few pulses.

The MBE limit corresponds to taking $\tau\to 0$ and $I\to 0$ with $I/\tau=F$, since this renders a constant flux $F$. In the MBE regime it was found that initially, when $N_1 \gg N$, the island density grows as $N\sim t^3$. However, when $N \gg N_1$, the growth of the island density becomes slower, crossing over to the power law $N\sim t^{1/3}$ (see Ref.~\cite{bara95}).

\begin{figure}
\begin{center}
\includegraphics[width=85mm]{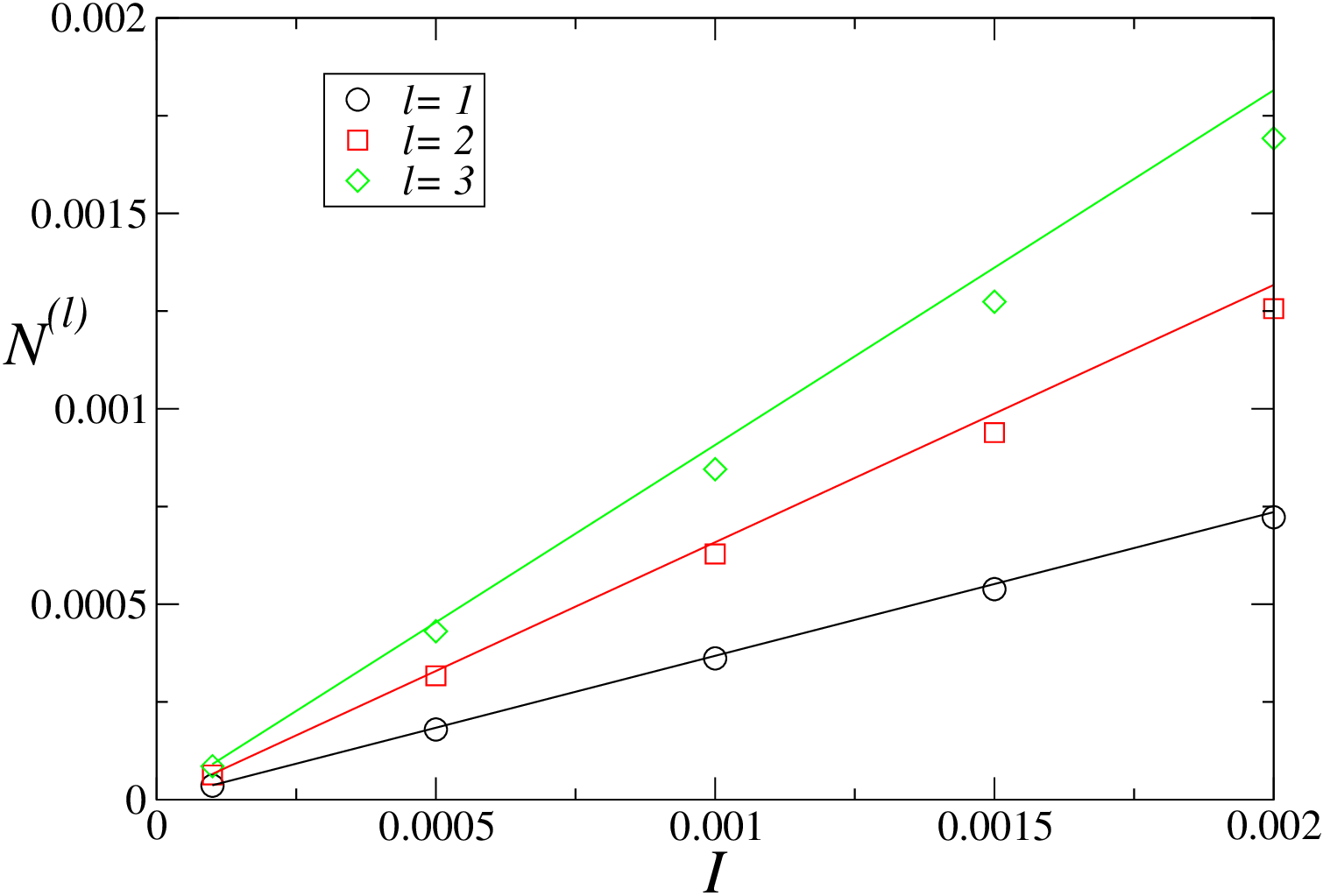}
\caption{(Color online) Island density in the PLD limit as a function of $I$ for the first three pulses $l= 1, 2, 3$ obtained from simulations (points) and from relations (\ref{eq1}), (\ref{eq2}), and (\ref{eq3}) (lines).
\label{fMF1a}}
\end{center}
\end{figure}

To solve the equations in the PLD limit, let us first consider the evolution between two pulses $l$ and $l+1$:
\begin{equation}
\frac{dN_1}{dt}= -2N_1^2- NN_1,
\label{eqpldn1}
\end{equation}
\begin{equation}
\frac{dN}{dt}= N_1^2,  
\label{eqpldn}
\end{equation}
where, without loss of generality, we set $D=1$. In order to make the above equations linear one introduces a modified time variable \cite{bril91, krap08}
\begin{equation}
T= \int_0^tN_1(t')dt'.
\end{equation}
Because of $dT=N_1(t)dt$, Eqs. (\ref{eqpldn1}) and (\ref{eqpldn}) turn into
\begin{equation}
\frac{dN_1}{dT}= -2N_1- N,
\end{equation}
\begin{equation}
\frac{dN}{dT}= N_1.  
\end{equation}
With the initial conditions $N_1(0)= I$ and $N(0)=N^{(l)}$, where $N^{(l)}$ is the island density after $l$ pulses, one is led to the solutions
\begin{equation}
N_1(T)= [I- (I+N^{(l)})T]\,\exp(-T),
\label{eqpldn1T}
\end{equation} 
\begin{equation}
N(T)= [N^{(l)}+ (I+N^{(l)})T]\,\exp(-T).
\label{eqpldnT}
\end{equation} 
In the PLD limit the pulses are so strongly separated in time that all diffusing adatoms have nucleated or attached to existing islands before the next pulse arrives. The diffusion process ends at the final (modified) time $T_f$ when 
\begin{equation}
N_1(T_f)= 0\,
\label{eqn1tf}
\end{equation}
and the corresponding island density  $N(T_f)= N^{(l+1)}$ stays constant until the next pulse arrives. Because of Eq.~(\ref{eqpldn1T}) the modified final time $T_f$ is given by $T_f=I/(I+N^{(l)})$, hence with Eq.~(\ref{eqpldnT}) we obtain the recurrence relation
\begin{equation}
N^{(l+1)}= (I+N^{(l)})\exp\bigg(\frac{-I}{I+N^{(l)}}\bigg),
\label{eqrec}
\end{equation}
with the initial condition $N^{(0)}=0$. Rewriting the island density as
\begin{equation}
N^{(l)}= A_lI
\label{eq1}
\end{equation}
with 
\begin{equation}
A_0= 0
\label{eq2}
\end{equation}
this recurrence relation turns into
\begin{equation}
A_{l+1}= (1+A_{l})\exp\bigg(\frac{-1}{1+A_{l}}\bigg),
\label{eq3}
\end{equation}
which is independent of $I$. This means that within this mean field approximation the island density is proportional to the pulse intensity. As shown in Fig.~\ref{fMF1a}, where relations (\ref{eq1}), (\ref{eq2}) and (\ref{eq3}) are compared with simulations, this is indeed the case for the first few pulses. For the predicted slope, however, we see a better agreement for the first pulse and also for smaller values of the pulse intensity. 

%================================================
%================================================

\begin{figure}
\begin{center}
\includegraphics[width=85mm]{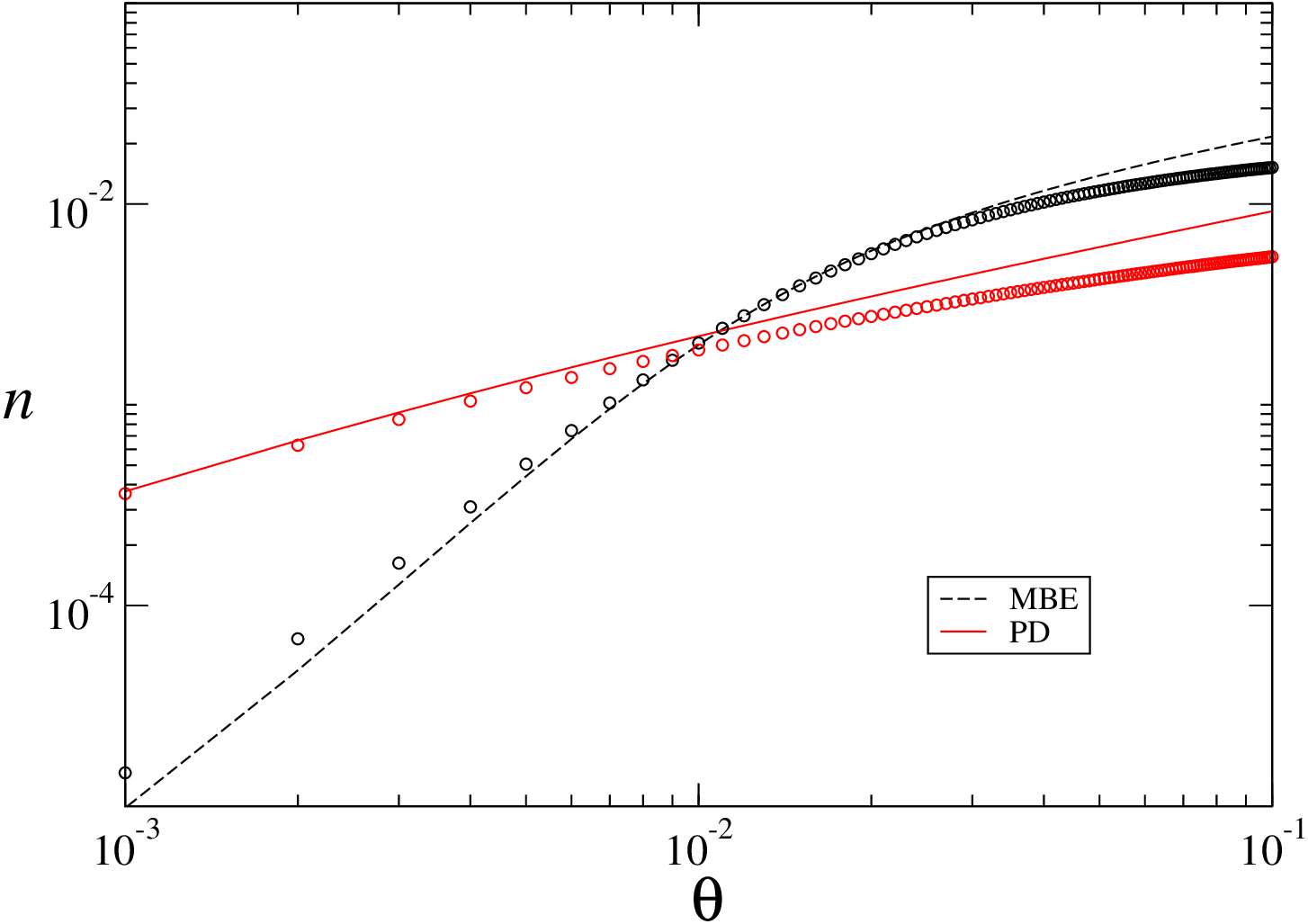}
\caption{(Color online) Comparison of the nucleation density obtained by simulations (points) and the rate equations (\ref{eqN1}) and (\ref{eqN}) (lines) with $I=0.001$ and, in the MBE limit, $R=10^4$. Note that for comparing the two cases for given $I$, $R$ has to be chosen such that  $R \ll I^{1/(2\gamma-1)}$.
\label{fMF1b}}
\end{center}
\end{figure}

This approximation reproduces another important feature of the full model, namely, after the first pulse, for a fixed value of $I$, the nucleation density in the PLD limit is considerably larger than the nucleation density in the MBE limit  $R \ll I^{1/(2\gamma-1)}$ (see Fig.~\ref{fMF1b}, where the rate equations and simulation results  are compared in both limits). However, the nucleation density in the MBE limit grows faster and eventually exceeds the nucleation density in the PLD limit such that the two curves in Fig.~\ref{fMF1b} cross each other. 

%================================================
%================================================

The rate equations used here assume islands to be point-like objects. Obviously, this approach works only for very small values of the coverage (much smaller than the value at which coalescence starts) and thus it is inadequate to fully account for the crossover from PLD to MBE. In particular, it does not allow one to confirm the scaling relation (\ref{eqcross}). In the next section we consider improved rate equations that overcome these problems. 

%-------------------------------------------------------
\section{Improved rate equations}
%-------------------------------------------------------

In order to improve the rate equations for the model defined in the introduction let us now consider the dynamics of islands with different sizes separately and study how they compete for the diffusing monomers and how other adatoms are deposited on top of these islands. 

A set of rate equations for MBE that takes these features into account is \cite{bale94}
\begin{eqnarray}
\frac{dN_s}{d\theta} \;=\; (D/F)N_1\left(\sigma_{s-1}N_{s-1}-\sigma_sN_s\right)\nonumber\\
+k_{s-1}N_{s-1}-k_sN_{s},
\end{eqnarray}
\begin{eqnarray}
\frac{dN_1}{d\theta} \;=\; 1- (D/F)N_1(2\sigma_{1}N_{1}+\sum_{s=2}^{\infty}\sigma_sN_s)\nonumber\\
-k_{1}N_{1}-\sum_{s=1}^{\infty}k_sN_{s},
\end{eqnarray}
where the constants $\sigma_{s}$ are effective rates describing how an island of size $s$ competes for the diffusing monomers while the constants $k_s$ are related to the capture of the incident monomers. These quantities are given by~\cite{bale94}
\begin{equation}
k_s= s^{2/d_f}
\end{equation} 
and 
\begin{equation}
\sigma_s= 2\pi(R_s/\xi)\frac{K_1(R_s/\xi)}{K_0(R_s/\xi)}\,,
\end{equation}
where $K_0$ and $K_1$ are modified Bessel functions, $R_s$ is the island radius, and~$\xi$ is the average distance a monomer travels before being captured by an island or another monomer, given by
\begin{equation}
\xi^{-2}= 2\sigma_1N_1+\sum_{s=2}^{\infty}\sigma_sN_s+(F/D)k_1N_1\,.
\end{equation} 
These improved relations assume that the islands are circular. However, in reality and simulations the islands resemble fractal objects. The fractal morphology of the islands is taken into account in an effective way by assuming that the island radius grows with its size by~\cite{bale94}
\begin{equation}
R_s= \alpha \, s^{1/d_f},
\label{raio}
\end{equation}  
where $\alpha$ is a fit parameter, to be determined by the comparison of the results arising from  numerical integration of the rate equations and simulations, and $d_f= 1.72$ is the fractal dimension of DLA-like clusters~\cite{tang93}.

In order to generalize the above rate equations to pulsed deposition we use, as in the former case, a discontinuous flux of incoming monomers, obtaining the differential equations
\begin{eqnarray}
\frac{dN_1}{d\theta}= \sum_l\bigg(1- k_1N_1- \sum_{s=1}^\infty k_sN_s\bigg)I\delta(\theta-lI)\nonumber\\
- 2D\sigma_1N_1^2 - DN_1\sum_{s=2}^{\infty}\sigma_sN_s,
\label{eqN1i}
\end{eqnarray}
\begin{eqnarray}
\frac{dN_s}{d\theta}= \sum_l[k_{s-1}N_{s-1}- k_sN_s]I\delta(\theta-lI)\nonumber\\
+ DN_1[\sigma_{s-1}N_{s-1}-\sigma_sN_s],
\label{eqNi}
\end{eqnarray}
where the $\delta$-functions couple to the terms proportional to $k_s$ since those terms are related to the capture of arriving monomers, an event that takes place during deposition. As in the previous case, the improved equations predict that the nucleation density $n$ and island density $N$ are identical since coalescence is not taken into account. Since coalescence influences directly only the island density, we find it more appropriate to compare the results coming from numerical integration of these equations with the nucleation density obtained by simulations.

%================================================
\begin{figure}
\begin{center}
\includegraphics[width=85mm]{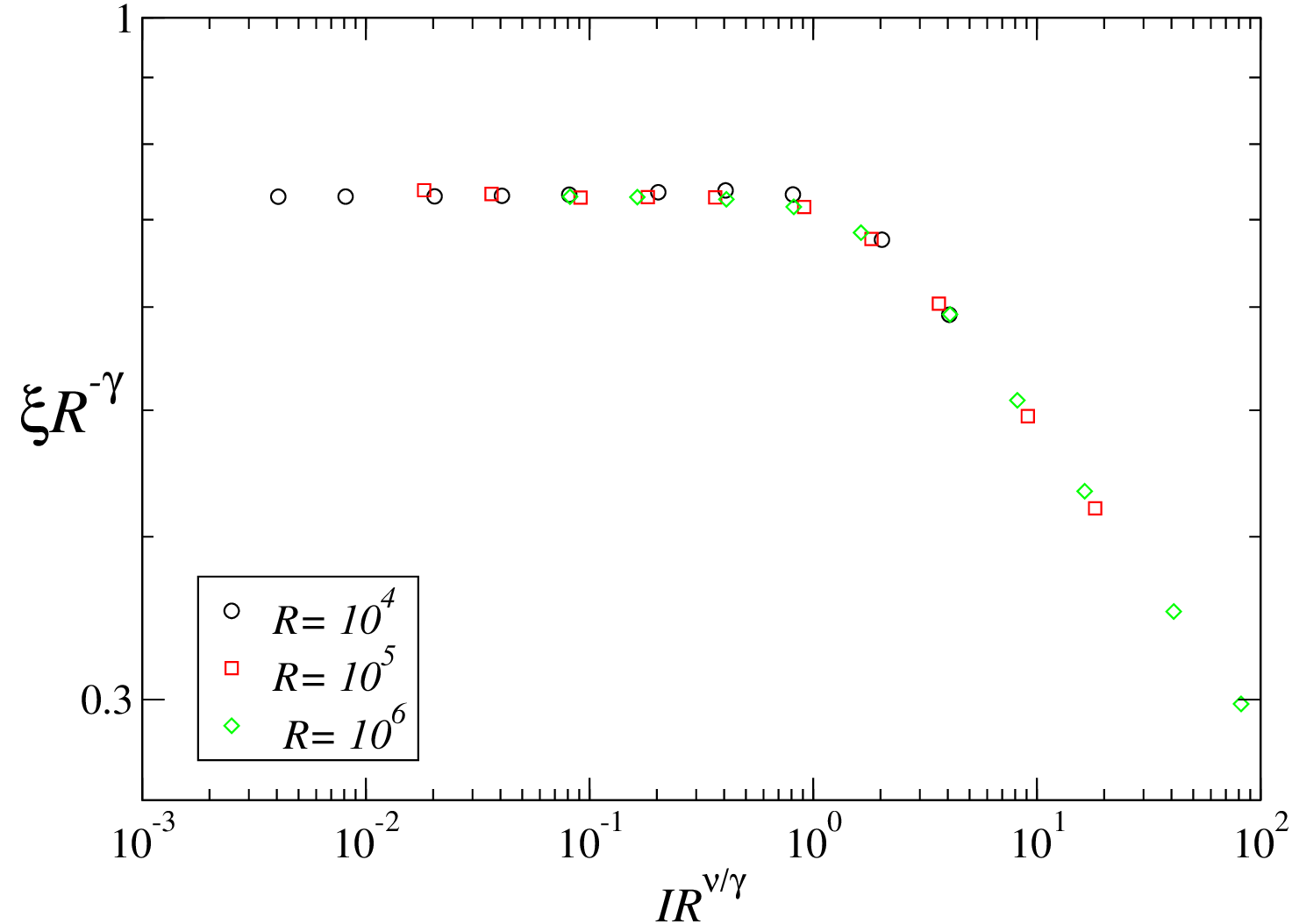}
\caption{(Color online) Data collapse to verify equation (\ref{eqcross}) obtained from numerical integration of (\ref{eqN1i}) and (\ref{eqNi}) with $\alpha=1$ at the coverage $\theta= 0.1$.
\label{fcross}}
\end{center}
\end{figure}
%================================================

The rate equations (\ref{eqN1i}) and (\ref{eqNi}) faithfully reproduce the predicted cross\-over from MBE to PLD. To demonstrate this we show, in Fig. \ref{fcross}, the corresponding data collapse according to the scaling form (\ref{eqcross}) with $\nu= 0.23$ and $\gamma= 0.15$, where we used $\alpha= 1$. We note that the value of the fit parameter $\alpha$ could be different for MBE and PLD. However, when considering the crossover from MBE to PLD one has to use a single value for which $\alpha=1$ turned out to be a good choice. This approximation is justified since the value of this parameter does not change the type of scaling behavior. Nevertheless, this ambiguity regarding the value of $\alpha$ may lead to small numerical deviations in the estimates of the scaling exponents $\nu$ and $\gamma$.

%================================================
\begin{figure}
\begin{center}
\includegraphics[width=85mm]{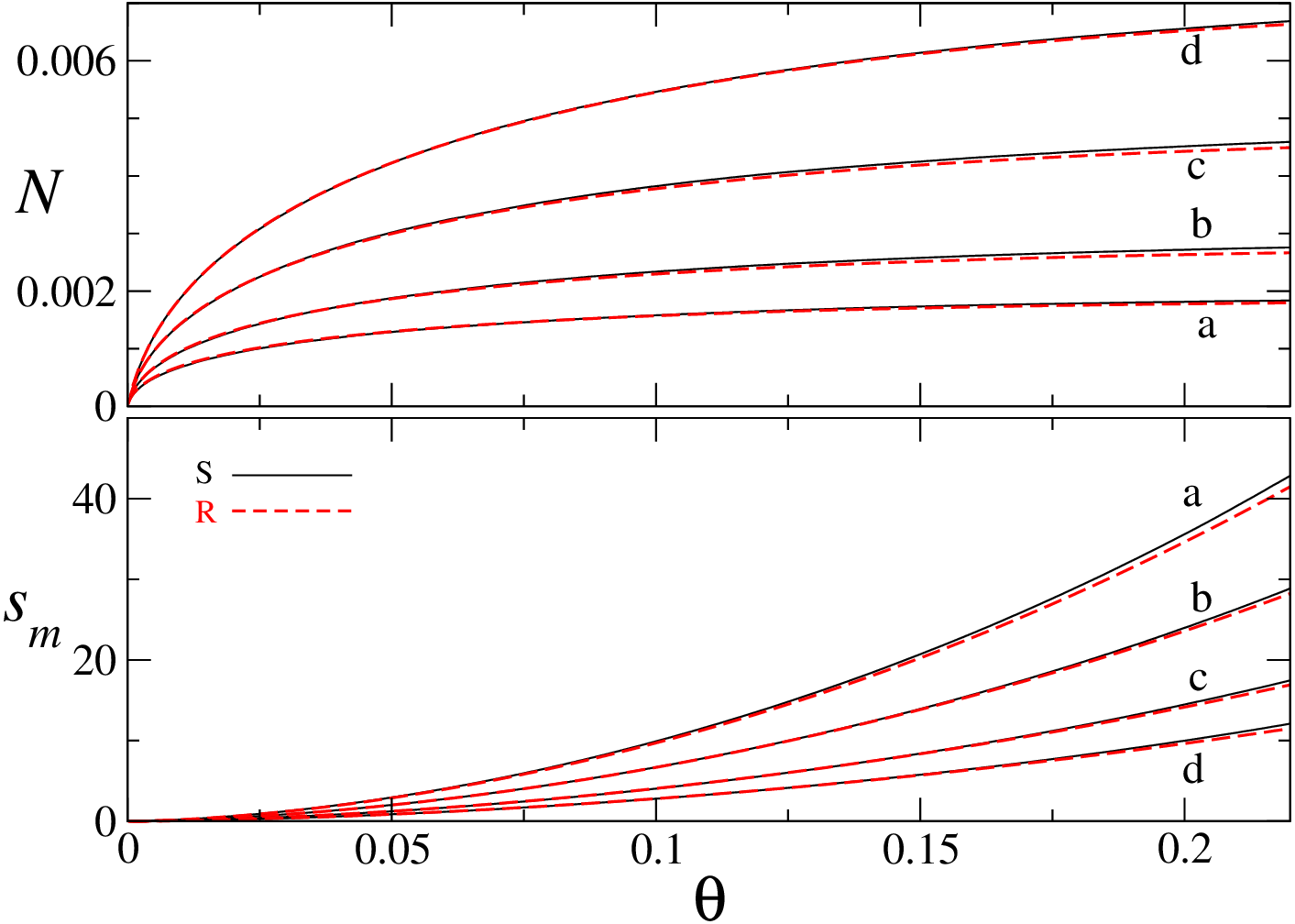}
\caption{(Color online) Comparison between simulations (S) and numerical integration of the rate equations (\ref{eqN1i}) and (\ref{eqNi}) (R) of the nucleation density (upper panel) and mean island size (lower panel) in the PLD limit for the pulse intensities $I=0.0001$ (a), $I=0.0002$ (b), $I=0.0005$ (c), and $I=0.001$ (d).
\label{fnuc}}
\end{center}
\end{figure}
%================================================

%================================================
\begin{figure}
\begin{center}
\includegraphics[width=85mm]{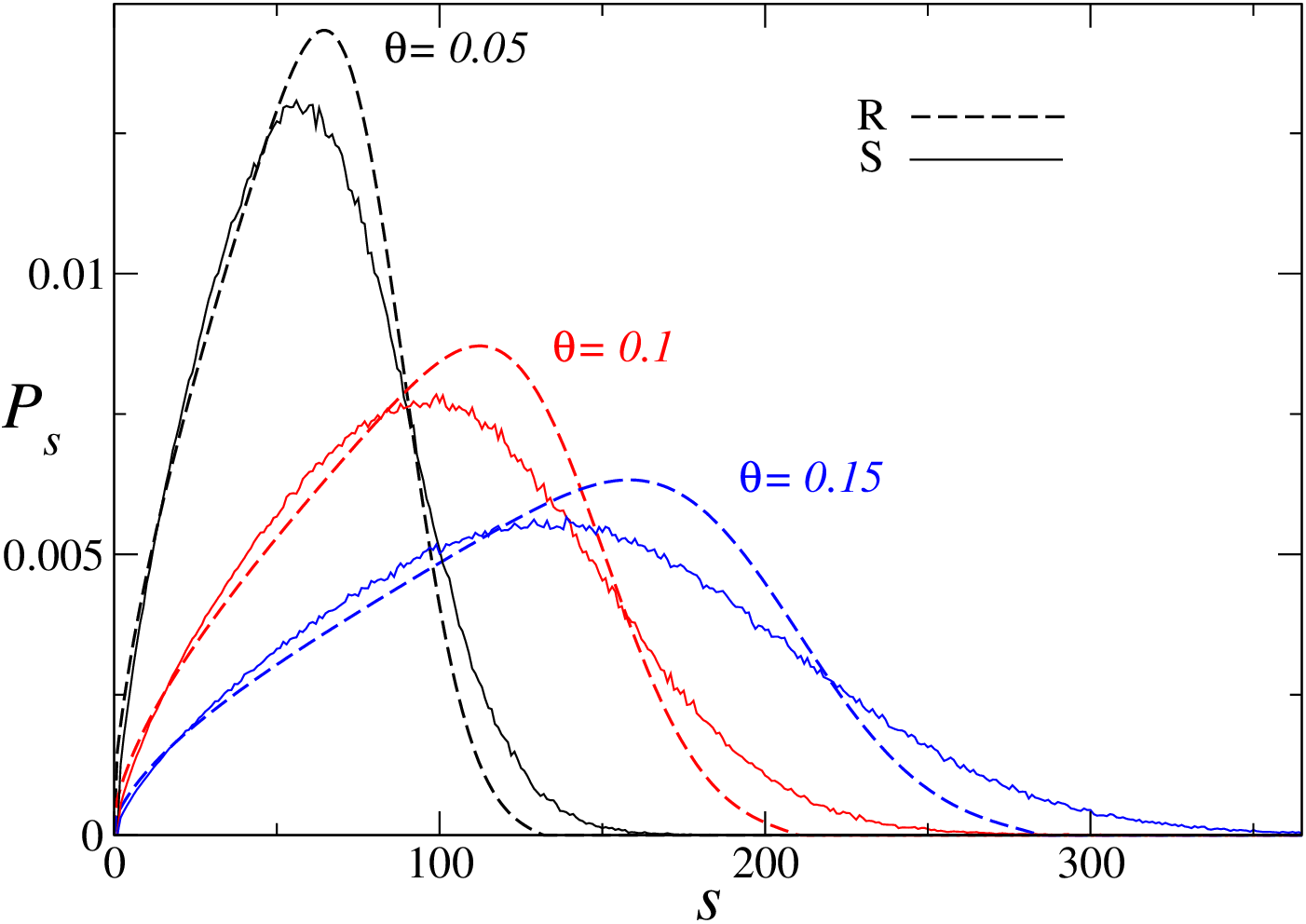}
\caption{(Color online) The probability distribution function (\ref{eqprob}) for $I= 0.0001$ in the PLD limit obtained from simulations (S) and from a numerical integration of the rate equations (R) (\ref{eqN1i}) and (\ref{eqNi}).
\label{fprob1}}
\end{center}
\end{figure}
%================================================

In Figs. \ref{fnuc} and \ref{fprob1} we compare the numerical integration of the rate equations with numerical simulations of the full model. Fig. \ref{fnuc} shows the nucleation density and the mean island size as functions of the coverage for different values of the pulse intensity. As can be seen, simulations and numerical integration of the rate equations agree almost perfectly. As is shown in Fig. \ref{fprob1}, the agreement is less good for the probability distribution (\ref{fprob1}). Such discrepancies are not surprising because the rate equations approach is a mean field theory, neglecting  correlations in the system. Similar discrepancies were already observed in the MBE limit~\cite{bale94}. The values of the fit parameter that we used are $\alpha\approx 1.7$ in the PLD limit and $\alpha\approx 0.3$ in the MBE limit.
  
This improved theory has considerable advantages when compared to the previous one. As shown in Fig.~\ref{fcross}, it fully accounts for the crossover from MBE to PLD, and the observables, except for coverages where coalescence starts to play an important role, are in excellent agreement with simulations. This excellent agreement indicates that the improved mean field equations capture the most important features of the model while other possible sources of deviations such as correlations and fluctuations are probably less important.

%-------------------------------------------------------
\section{Scaling properties of pulsed deposition}
%-------------------------------------------------------

As mentioned in the introduction, it was proposed that PLD could be described by an unconventional logarithmic scaling form~\cite{hinn01}. In this section we study this conjecture from a critical perspective.

Let us first recall the scaling properties of MBE. After an initial transient time and before the onset of coalescence, the probability distribution (\ref{eqprob}) is known to obey the scaling relation \cite{bale94}  
\begin{equation}
p_s(\theta)= \langle s\rangle^{-1}f\bigg(\frac{s}{\langle s\rangle}\bigg),
\label{eqscaprob}
\end{equation}
where $f(x)$ is a universal function independent of the control parameters $R$ and $\theta$. This means that, in the MBE regime, the model has only one characteristic size and therefore the morphology of the islands stay the same if the control parameters are varied.

\begin{figure}
\begin{center}
\includegraphics[width=85mm]{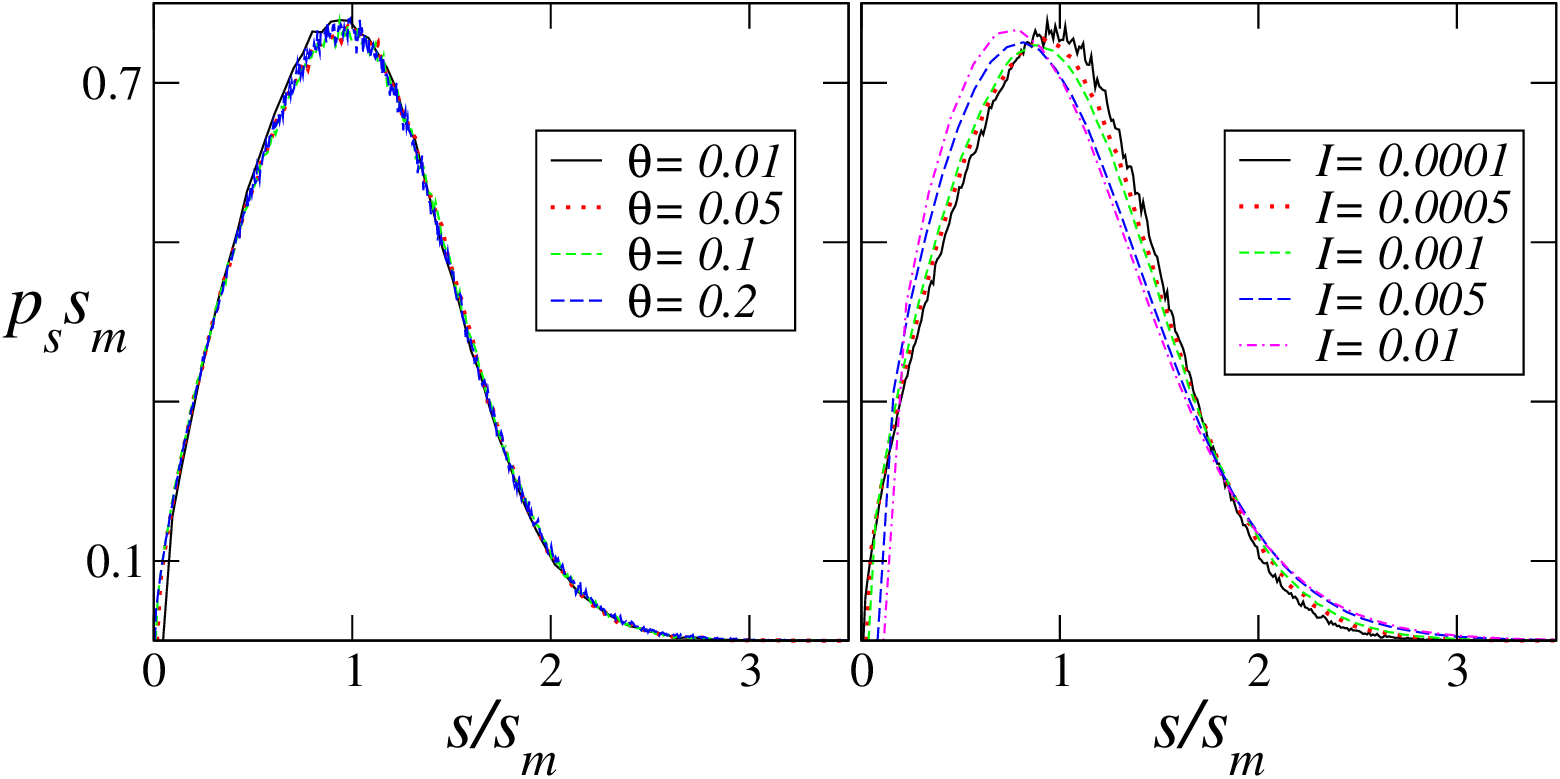}
\caption{(Color online) The probability distribution $p_s$, obtained from simulations, multiplied by mean island size $\langle s\rangle$ as a function of $s/\langle s\rangle$ for $I=0.0001$ and different values of $\theta$ (left); for $\theta= 0.1$ and different values of $I$ (right).
\label{fscaprob}}
\end{center}
\end{figure}

Turning to the PLD limit, in Fig. \ref{fscaprob} we observe that a convincing data collapse is obtained for different coverages if the pulse intensity is kept fixed, while the collapse is not satisfactory for different values of the pulse intensity and a fixed coverage. Therefore, in the PLD limit, the function $f(x)$ is not universal, instead it explicitly depends on one of the control parameters, namely, the pulse intensity $I$. However, for reasons that are explained below, the island morphology seems to be the same for different values of the pulse intensity.  

%================================================
%================================================

\begin{figure}
\begin{center}
\includegraphics[width=85mm]{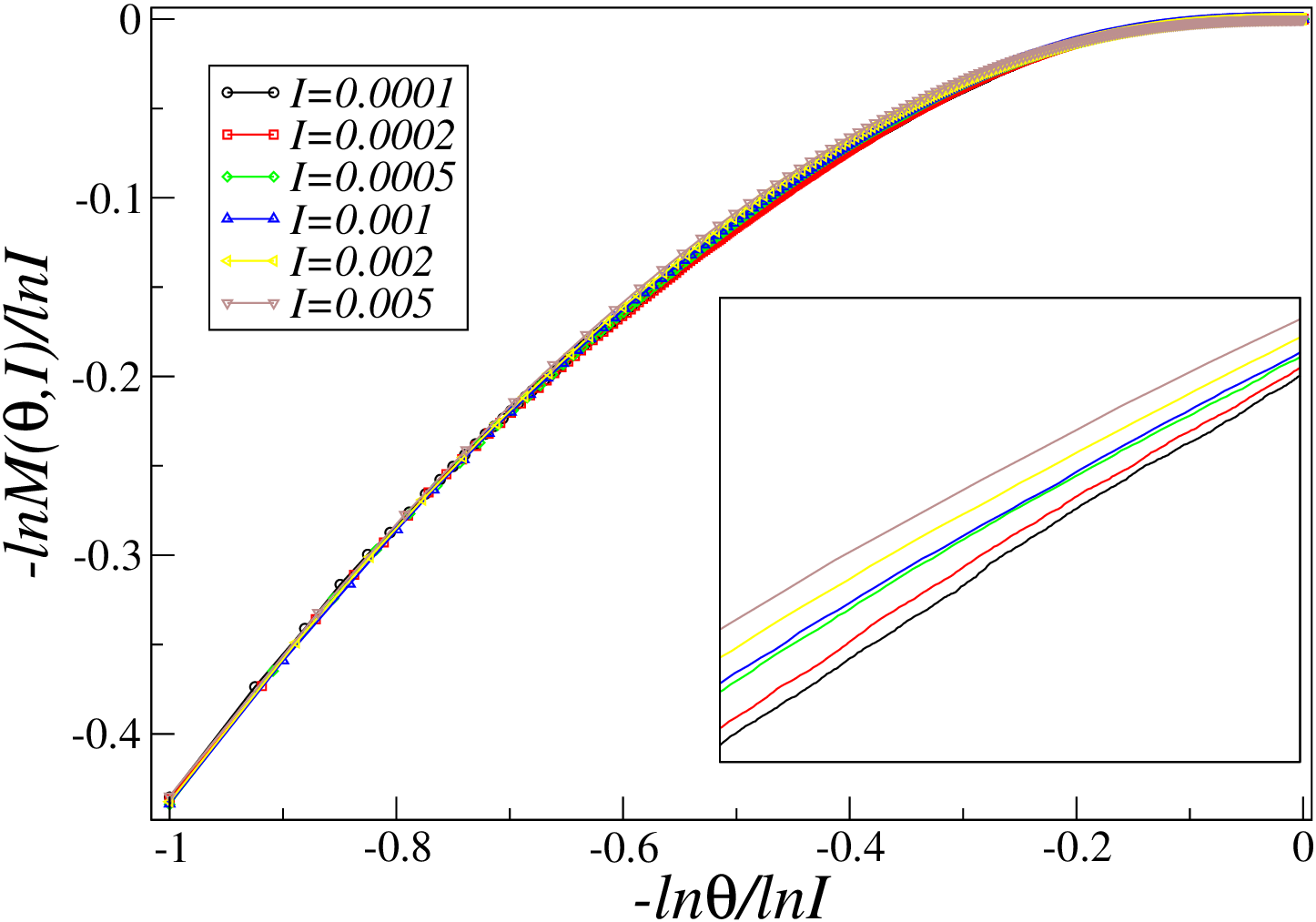}
\caption{(Color online) Data collapse of the normalized nucleation density, coming from simulations data, with the logarithmic scaling form. The inset shows the interval $0.2<\ln\theta/\ln I<0.25$ where the scaling function increases with the pulse intensity $I$.
\label{fsca1}}
\end{center}
\end{figure}

To explain the failure of conventional scaling, the logarithmic scaling form
(\ref{eqscalog}) was suggested on a purely heuristical basis~\cite{hinn01}. This scaling form had been applied previously to various other problems: experiments in turbulence \cite{meakb86}, self-organized critical sandpile models \cite{kada89, teba99}, and, as in the present case, DLA-related growth processes \cite{wu90}. Fig.~\ref{fsca1} shows a data collapse of the normalized nucleation density according to this scaling form, which at first glance seems to be convincing. Initially it was speculated that the unusual type of scaling behavior, which can be explained in terms of continuously varying critical exponents~\cite{sitt02}, may be related to the fractal structure of the islands, which become more and more compact as the first monolayer is filled up. However, later it was shown~\cite{lee03} that the same logarithmic scaling form can be used in a 1+1-dimensional model for pulsed deposition, where the (one-dimensional) islands are always compact. 

A closer look at Fig.~\ref{fsca1} reveals that the collapse is not perfect, rather there are deviations with a clear systematic tendency. It seems that the curves become more straight as the intensity is reduced, questioning the concept of logarithmic scaling.

%========================= AN ALTERNATIVE SUGGESTION =====================

Although we are still unable to disprove or confirm the concept of logarithmic scaling applied to PLD, it is in our opinion useful to demonstrate that an alternative scaling concept yields at least as good if not even better results. Starting point is the observation that average quantities like the mean island size and the nucleation density exhibit a clean power-law dependence on $I$ if the coverage $\theta$ is kept fixed. However, the value of the exponent varies with the coverage, i.e.
\begin{equation}
n(I,\theta)\sim I^{\alpha(\theta)}.
\label{eqnlog}
\end{equation} 
%
%

%================================================
%================================================

%================================================
\begin{figure}
\begin{center}
\includegraphics[width=85mm]{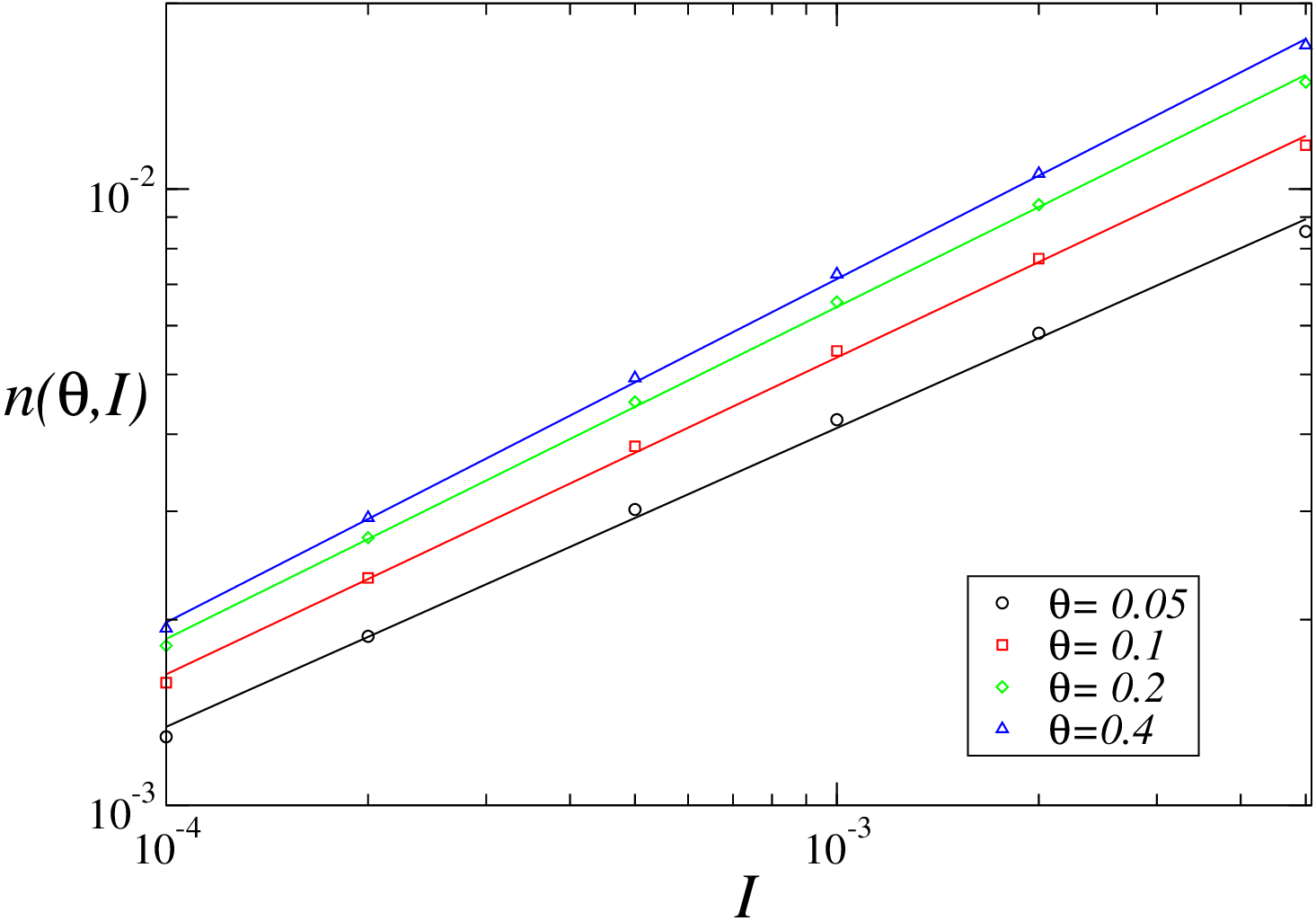}
\caption{(Color online) Nucleation density, obtained from simulations, as a function of $I$ for different values of $\theta$. The graph suggests that the exponent $\alpha$ increases with $\theta$.
\label{fnXI}}
\end{center}
\end{figure}
%================================================

\noindent
The plots of the nucleation density as a function of $I$ for different values of $\theta$ (cf. Fig. \ref{fnXI}) suggest that the exponent $\alpha$ increases monotonically with $\theta$. Actually, $\alpha= 2\nu$ and therefore Eq.~(\ref{eqldp}) should be corrected by introducing an exponent $\nu$ which depends continuously on the coverage. Since the exponent~$\nu$ is related to the fractal dimensionality of the islands, this indicates that the fractal dimension of them effectively varies with $\theta$. However, the data collapse of the probability distribution for different values of the coverage (Fig. \ref{fscaprob}) shows that this is clearly not the case.

We have also observed that in the one dimensional model, where the islands have no fractal properties, the nucleation density follows a power law for a fixed value of the coverage. Moreover, we observed that the function $f(x)$, defined in Eq.~(\ref{eqscaprob}), does depend on the pulse intensity. This indicates that these two features of the model in the PLD limit are not related to the fractal dimension of the islands and hence we do not expect it to vary with the control parameters.

In Fig.~\ref{fsca1} we present a data collapse using relation~(\ref{eqscalog}) for different values of the pulse intensity. Since $M(I,1)=1$ and $M(I,I)$ follows a power-law, the first ($\theta= I$) and last ($\theta= 1$) points of the curves collapse. However, the situation in the middle of the curve ($I\ll \theta \ll 1$) is unclear. More specifically, as is shown in the inset of Fig. \ref{fsca1}, for $0.2<\ln\theta/\ln I<0.25$ the value of $\ln M(I,\theta)$ increases \textit{monotonically} with $I$, thus exhibiting a systematic deviation rather than a statistical error. Also, this scaling is in disagreement with relation~(\ref{eqnlog}), since from it the value of $M(I,\theta)$ should be constant for a fixed coverage. For these reasons we consider the numerical evidence supporting equation (\ref{eqnlog}) as more reliable than the data collapse obtained using the logarithmic scaling form.

%-------------------------------------------------------
\section{Conclusion}
%-------------------------------------------------------

In this paper we have studied several variants of rate equations for pulsed deposition. First a simple set of rate equations, where the islands are treated as point-like objects, was considered. The equations were solved exactly in the limit of strong and temporally separate pulses (the so-called PLD limit as opposed to the MBE limit of continuous deposition), reproducing some features of the model for very small values of the coverage. Since this approximation does not take the dimension of the islands into account it was not possible to faithfully reproduce the crossover from MBE to PLD. This problem was overcome with a second improved set of rate equations that takes the dimension of the islands into account. As in the case of MBE \cite{bale94}, we showed that these improved equations lead to results which are in excellent agreement with simulations for pulsed deposition.

Another point of the present work was to revisit a recently proposed logarithmic scaling for pulsed deposition from a critical perspective. In the corresponding data collapses we have observed small violations for intermediate coverages with a systematic drift, indicating that logarithmic scaling may be a good approximation for pulse intensities in computer simulations and experiments but probably it is not asymptotically valid in the limit $I\to 0$.

As an alternative suggestion, we have proposed that the nucleation density, for a fixed value of the coverage, as a function of $I$ follows a power law and that the exponent varies with the coverage. This suggestion is supported by numerical simulations and leads to numerical results which are more accurate than the data collapses obtained by using logarithmic scaling. This may be another hint that logarithmic scaling, as proposed in Ref.~\cite{hinn01}, has to be replaced by a different type of scaling theory.

We have pointed out another new feature of the two-dimensional model in the PLD limit, namely, the probability distribution $p_s$ that a site belongs to a cluster of size $s$ does not scale according to relation (\ref{eqscaprob}) for different values of the pulse intensity. At first glance this non-universal behavior may lead to the conclusion that for different pulse intensities the fractal dimension of the islands is different. However, studying the one-dimensional model, where islands have no fractal properties, the same effect is observed, indicating that this is not the case.

Let us finally comment on the one-dimensional model in more detail. This case was previously studied in Ref.~\cite{lee03}, where a reasonable data collapse for the nucleation density based on logarithmic scaling was presented. Contrarily, we observe a clean power law behavior of the nucleation density as a function of $I$ for a fixed $\theta$ and also a non-universal probability distribution for different values of $I$. Hence, the one dimensional model seems to exhibit a behavior which is as rich as in the two-dimensional case and a systematic study of the 1d case may be very useful for clarifying some aspects of PLD.\\

\noindent \textbf{Acknowledgments:}\\
We are grateful to P. L. Krapivsky for sending us unpublished notes on pulsed deposition, which influenced the present work. Moreover, we thank S. R. Dahmen for critical reading of the manuscript. This work was supported financially by the Deutsche Forschungsgemeinschaft (grant HI 744/3-1).


\begin{thebibliography}{99}
\bibitem{footnote}
When diffusion of atoms along the edges of islands is also permitted the islands can become compact objects ($d_f=2$)~\cite{lam02, bale94}.
\end{thebibliography}
\end{document}